# Intensity adaptive optics

Zimo Zhao[1,†], Yifei Ma[1,†], Zipei Song[1,†], Jacopo Antonello[1], Jiahe Cui[1], Binguo Chen[2], Jingyu Wang[1], Bangshan Sun[1], Honghui He[2], Lin Luo[3], Julian A.J. Fells[1], Steve J. Elston[1], Martin J. Booth[1], Stephen M. Morris[1], and Chao He[1,*]

[1]*Department of Engineering Science, University of Oxford, Parks Road, Oxford, OX1 3PJ, UK*
[2]*Guangdong Engineering Center of Polarization Imaging and Sensing Technology, Tsinghua Shenzhen International Graduate School, Tsinghua University, Shenzhen 518055, China*
[3]*College of Engineering, Peking University, Beijing 100871, China*
[†]*These authors contributed equally to this work*
[*]*Corresponding author: chao.he@eng.ox.ac.uk*

**Adaptive optics (AO) is a powerful tool employed across various research fields, from aerospace to microscopy. Traditionally, AO has focused on correcting optical phase aberrations, with recent advances extending to polarisation compensation. However, intensity errors are also prevalent in optical systems, yet effective correction methods are still in their infancy. Here, we introduce a novel AO approach, termed intensity adaptive optics (I-AO), which employs a dual-feedback loop mechanism to first address non-uniform intensity distribution and subsequently compensate for energy loss at the pupil plane. We demonstrate that I-AO can operate in both sensor-based and sensorless formats and validate its feasibility by quantitatively analysing the focus quality of an aberrated system. This technique expands the AO toolkit, paving the way for next-generation AO technology.**

## Introduction

Optical aberrations, including phase and polarisation, compromise the performance of optical systems[1-4]. For decades, adaptive optics (AO) has been a crucial technology in correcting phase aberrations, enhancing the performance of astronomical telescopes[5-7], optical communications[8-10], and super-resolution microscopy/nanoscopy[11-17]. Recently, AO has expanded into the vectorial domain, addressing either polarisation aberration alone[18-22] or the combination of polarisation and phase aberrations[23-25]. However, intensity errors remain underdeveloped in terms of available correction techniques, despite their adverse impact on system performance. In microscopy, for example, these errors can result from factors such as beam propagation[26,27], diattenuation effects[28], or absorption by materials or biological tissues[29-32], leading to degraded image resolution by disturbing the intensity distribution across the pupil plane of the objective lens.

To address these errors, we introduce intensity adaptive optics (I-AO), a new AO tool focused on achieving intensity uniformity at the pupil plane, with absolute intensity recovery as a secondary feature through a feedback mechanism. I-AO supports either direct intensity measurements at the pupil plane (sensor-based) or indirect intensity estimation from the focal region (sensorless). Figures 1(a) and 1(b) illustrate the optical system with spatially varying intensity losses (with Fig. 1(a) showing the three sources used throughout this work) and demonstrate its performance before and after integrating the I-AO correction module. Note that conventional AO primarily corrects phase aberrations between the pupil and focal plane but does not address spatially varying loss on a pixel-by-pixel basis at the pupil plane. Consequently, non-uniform intensity on this plane causes distorted focal spots due to imperfect interference which remains uncorrected by phase-only AO. Throughout this work, we classify spatially varying intensity loss at the pupil plane as intensity errors.

At the core of our I-AO methods, which distinguish them from existing AO techniques, is a dual-loop feedback mechanism (see Fig. 1(c) and Fig. 1(d) for schematics of the sensor-based and sensorless I-AO methods, respectively). The primary loop ensures a uniform intensity distribution at the pupil plane: in the sensor-based method, the intensity profile of the pupil plane is directly measured by a camera for subsequent compensation, whereas in the sensorless method, intensity error is estimated based on a sequence of images taken from the focal plane. The I-AO corrector (see Fig. 1(b)) then undertakes the task of restoring intensity uniformity. The secondary loop, identical in both methods, compensates for total energy loss by adjusting the attenuator following the light source. Detailed procedures for both methods are provided later.

In this work, we demonstrate the feasibility of the proposed I-AO methods by enhancing the quality of the focus in an aberrated focusing system. While I-AO has broad applicability, we first confine our discussion to microscopy, which is particularly beneficial for applications that highly demand a uniform pupil plane intensity and precise absolute intensity levels, such as pathology and clinical diagnostics[33].

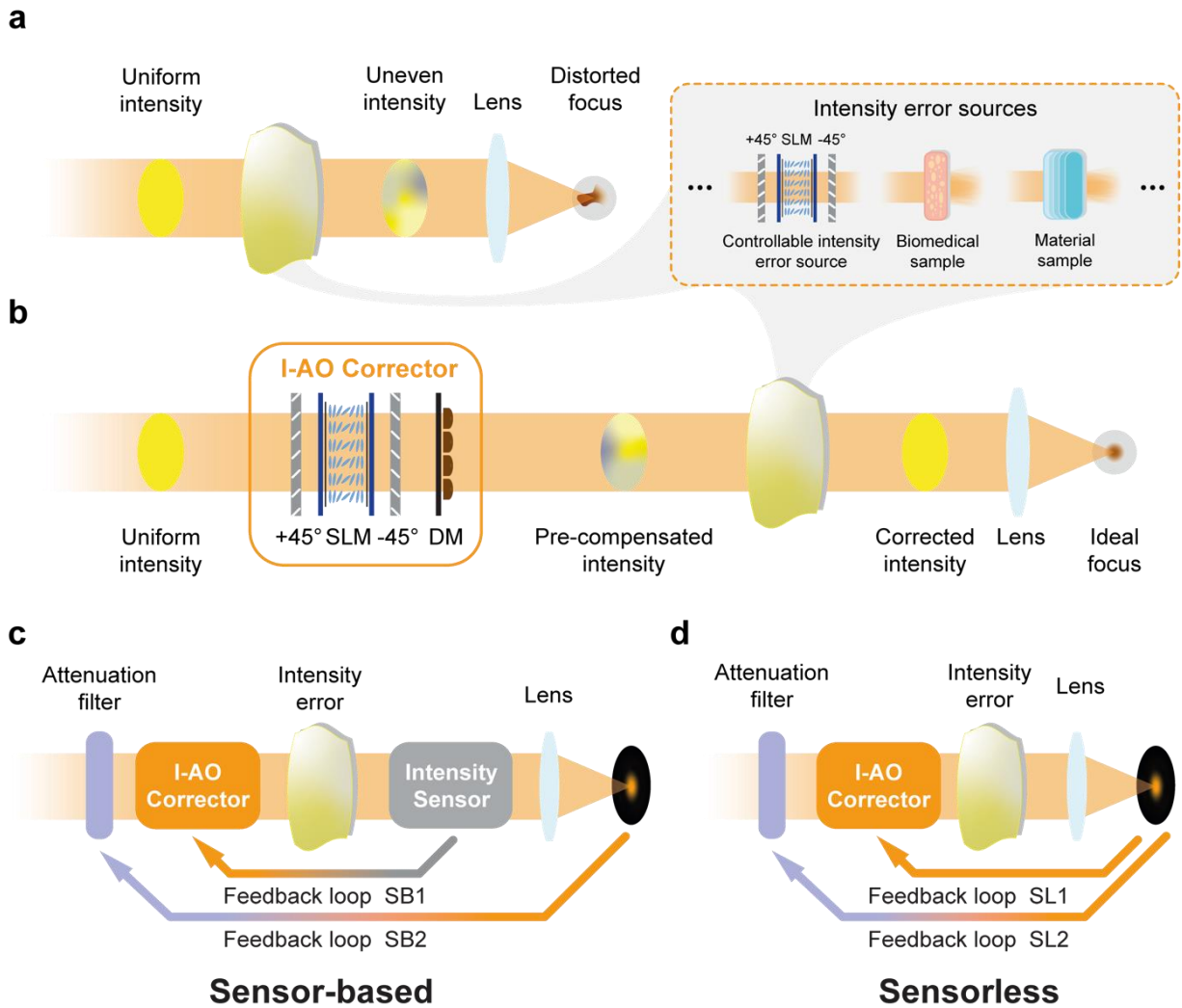

**Figure 1: The concept of I-AO.** (a) An optical system affected by spatial intensity error sources, with a distorted focal spot. In this work, spatially varying intensity losses are introduced through three sources: 1) a spatial light modulator (SLM) sandwiched by two crossed polarisers for precise control of intensity distribution, generating arbitrary intensity errors; 2) a thin biomedical sample (fibrotic tissue); and 3) a material sample (birefringent crystals). (b) Integration of an I-AO corrector restores intensity uniformity, addressing spatially varying loss to achieve an ideal focus. The I-AO corrector employs an SLM sandwiched by two crossed polarisers for pixelated intensity correction at the pupil plane, with an additional deformable mirror (DM) to compensate for phase errors. (c) Conceptual schematic of sensor-based I-AO with dual-loop feedback correction (SB1 and SB2), where an intensity sensor measures the intensity errors directly from the pupil plane. (d) Conceptual schematic of sensorless I-AO with dual-loop feedback correction (SL1 and SL2), where the intensity error is estimated using images obtained from the focal plane. The experimental setup is shown in **Supplementary Note 1**, and details of the dual-loop mechanism can be found in **Supplementary Note 2**.

## Results

For both I-AO methods (sensor-based and sensorless), two key points should be noted: 1) both methods simultaneously correct phase and intensity errors, addressing not only systematic phase errors but also additional phase errors induced by the AO device itself (due to geometrical effects) alongside intensity errors; 2) both methods go beyond the single-loop feedback mechanism of conventional AO, employing a dual-loop approach. Extending existing AO methods to this intensity analogue is a non-trivial task because intensity is related to energy. This fundamental difference means that the modal concept in traditional sensorless phase AO cannot be directly applied to sensorless I-AO, thus requiring novel developments. We now elaborate on each method in detail.

**Sensor-based I-AO**

The sensor-based I-AO approach involves the following pipeline: 1) A reference profile is first acquired by capturing the full intensity distribution at the pupil plane without introducing intensity errors, using an intensity sensor such as a monochrome camera. 2) A distorted intensity profile is then acquired at a plane corresponding to the conjugate pupil plane of the objective lens within the aberrated imaging system, providing feedback data for the I-AO corrector. 3) Next, using the obtained two profiles, the required retardance values for intensity pre-compensation are calculated at the conjugate plane of the I-AO corrector to initiate the first feedback correction loop (SB1 in Fig. 1(c)). 4) These calculated retardance values are then applied to the I-AO corrector to achieve uniform intensity across the pupil. 5) The DM is further used to perform sensorless phase AO and compensate for both geometrical phase errors induced by the SLM itself and residual systematic phase errors. This completes the first feedback loop. 6) Finally, with a uniform intensity profile obtained at the focal plane, the attenuator following the light source is adjusted to match the total intensity at the pupil plane to the level recorded in the reference frame. This completes the second feedback loop (SB2 in Fig. 1(c)). Details on the SLM calibration procedure and its application in I-AO are provided in **Supplementary Note 3**. The sensor-based dual-loop feedback correction is further elaborated in **Supplementary Note 4**.

This methodology was validated through two experiments to demonstrate its necessity and feasibility. The first experiment underscored the need for I-AO. Initially, we introduced an external, spatial intensity error into the system (see Fig. 2(a), I-AO off) using an additional SLM positioned between two polarisers (see **Supplementary Note 1**) and attempted to correct it with sensorless phase AO via the DM (see Fig. 2(a), Phase AO on). The results showed that it is still an aberrated focal spot, indicating that conventional phase AO correction alone cannot fully rectify the effects of intensity errors (also see **Supplementary Note 5**). After removing the introduced intensity error, the original distribution was restored across the pupil to achieve a diffraction-limited focal spot. Figure. 2(b) shows the intensity values across a cross-section of the focal spot at each step. This validation confirms that I-AO is essential for completing the correction of intensity errors.

In the second experiment, following the previously outlined pipeline, we performed sensor-based I-AO correction for the same intensity error as in the first experiment, with the pupil and focal plane profiles shown in Fig. 2(c) (I-AO off). In the first feedback loop SB1, intensity non-uniformity was corrected based on the measured pupil image, resulting in a uniform intensity profile across the pupil as shown in Fig. 2(c) (I-AO SB1). Subsequently, sensorless phase AO was performed (see Fig. 2(c), Phase AO on) to complete the first feedback loop SB1. Comparing the intensity profiles before and after I-AO correction reveals enhanced pupil intensity uniformity, along with significant improvement in focal spot quality. Finally, feedback loop SB2 was implemented (see Fig. 2(c), I-AO SB2) to enhance the total intensity. The corresponding focal spot cross-sections for each step are illustrated in Fig. 2(d).

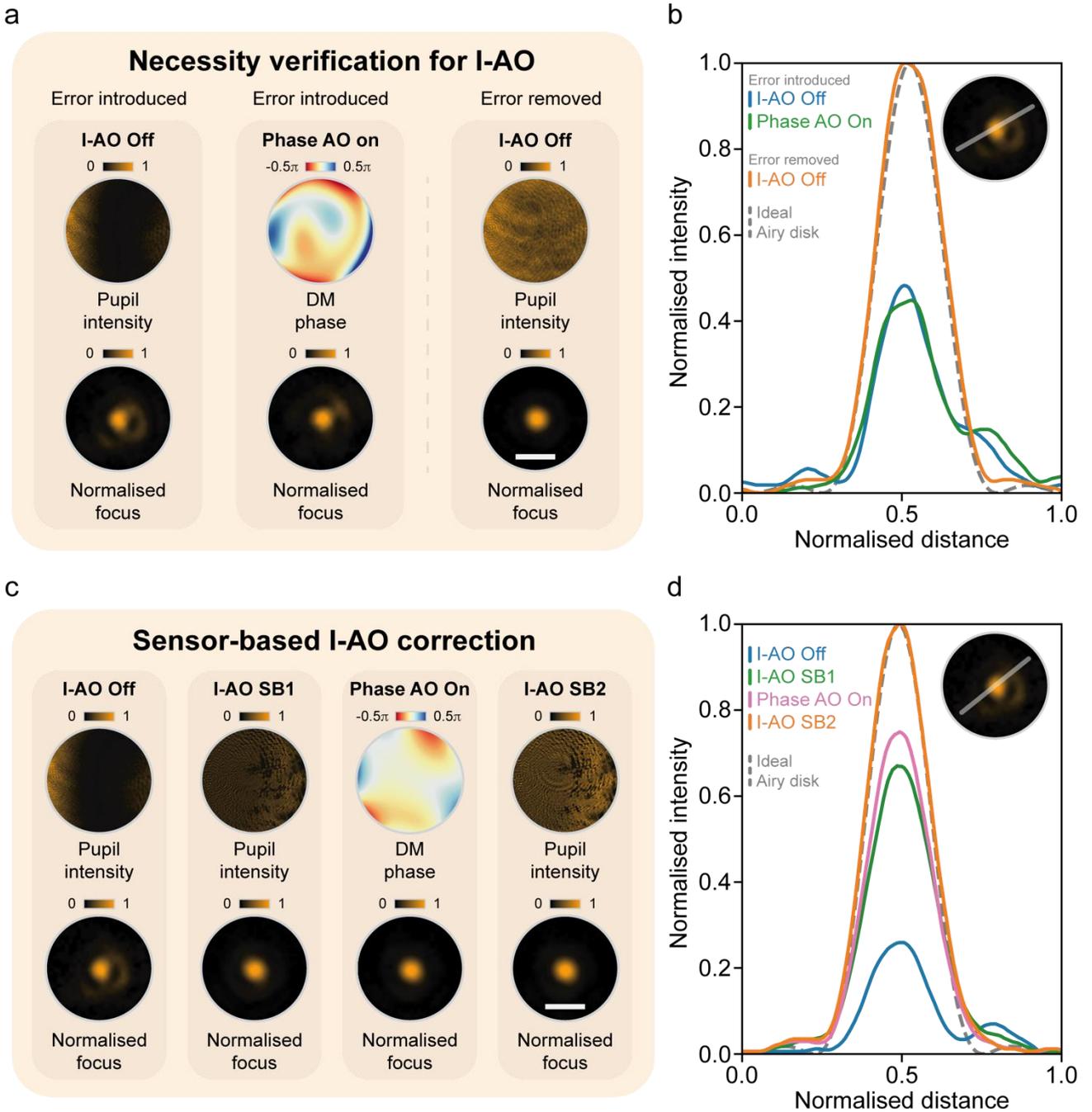

**Figure 2: Sensor-based I-AO for intensity error correction.** (a) Demonstration of the necessity of I-AO for intensity error correction by comparison with the results of traditional phase AO correction. The pupil intensity, DM phase profile, and normalised focus profiles are shown under three conditions: I-AO off with introduced intensity error, phase AO on, and I-AO off without introduced intensity error. (b) Cross-sectional intensity profiles of the focus from (a), with the ideal Airy disk indicated by a dashed line. (c) Sensor-based I-AO correction using a dual feedback loop (SB1 and SB2 in Fig. 1(c)). The pupil intensity, DM phase profile, and normalised focus profiles are shown for each step of the correction process. (d) Cross-sectional intensity profiles of the focus from (c), compared with the ideal Airy disk. Note that in (c), the final step (I-AO SB2) increases the light source intensity, leading to a higher final intensity compared to earlier steps and scaling other curves after the normalisation in (d). White scale bars in (a) and (c) represent 30 μm for all normalised focus profiles.

The experiments demonstrate that the sensor-based I-AO method significantly improved the performance of an intensity-aberrated optical system, with a primary focus on enhancing intensity uniformity at the pupil plane and a complementary improvement in total intensity.

**Sensorless I-AO**

We next demonstrate the feasibility of the sensorless I-AO approach, which eliminates the need for direct sensor-based measurement of intensity errors at the pupil plane. This sensorless approach is particularly useful for compact optical systems where complex sensors cannot be used. Like conventional phase AO, sensorless I-AO sequentially applies a series of pre-designed correction patterns (modes) to the AO device, indirectly inferring intensity errors by analysing the focal spot image as feedback. The optimal correction pattern is selected and applied to the corrector by maximising image quality metrics.

However, new modes must be specifically designed because: 1) intensity is an energy related quantity, which makes it unsuitable to use conventional orthogonal Zernike modes as the basis for correction; 2) similar to the sensorless techniques in modern vectorial AO correction methods, the I-AO approach must address the complex interplay between intensity and phase due to geometrical effects when modulating the SLM, requiring simultaneous correction of any additional phase errors introduced by the SLM. To address these challenges, we developed a new set of correction modes – termed intensity Zernike modes; further details are provided in **Supplementary Note 6**.

The following steps outline the correction process: 1) First, 15 intensity Zernike modes are evaluated and analysed to initiate the first feedback correction loop (SL1 in Fig. 1(d)), including four standard correction modes – piston, tip, tilt, and defocus – that are typically not corrected in conventional phase AO; 2) Images of the focal spot are then recorded as each intensity Zernike mode is applied at varying amplitudes. Unlike conventional phase AO, which applies a combination of Zernike modes during aberration correction[34], each correction mode in I-AO is applied individually with its corresponding coefficient; 3) Next, a novel approach combining two image metrics – low spatial frequency content of the captured image[35] and spot circularity – is employed to achieve the best focus profile, ultimately determining the optimal intensity Zernike mode and coefficient for correction. This completes the first feedback loop. 4) Finally, after obtaining a uniform intensity profile at the focal plane, the attenuator is adjusted to recover the total intensity level. This ends the second feedback loop (SL2 in Fig. 1(d)). See **Supplementary Note 7** for a detailed mechanism.

Following this, we conducted experiments to validate the sensorless I-AO approach, using the same external intensity error as in the sensor-based I-AO validation. Fig. 3(a) presents the stepwise results of the correction process. Initially, the intensity error creates an aberrated focal spot (I-AO off). Sensorless phase AO is then applied to address the residual phase errors in the system, with the applied correction phase pattern and resulting focal spot shown in Fig. 3(a) (Phase AO on). Next, the feedback loop SL1 identifies the optimal intensity Zernike mode and coefficient, which results in the restored focal spot (I-AO SL1). A subsequent phase AO correction refines intensity uniformity, producing the final DM phase pattern and the corresponding focal spot (Phase AO SL1 in Fig. 3(a)). This is followed by the feedback loop SL2, which compensates for total intensity loss (I-AO SL2). Cross-sectional intensity profiles in Fig. 3(b) illustrate the progressive improvement in uniformity and shape throughout the correction process.

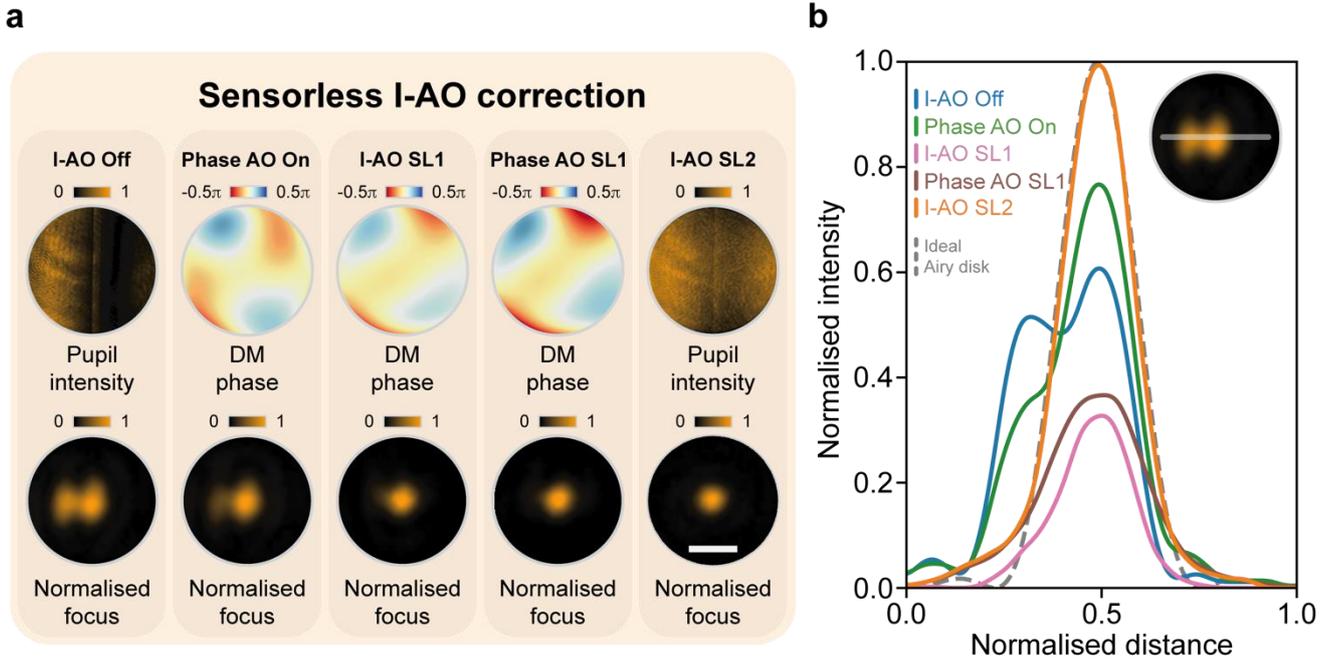

**Figure 3: Sensorless I-AO for intensity error correction. (a)** Sensorless I-AO correction procedure. The pupil intensity profiles, DM phase profiles, and normalised focal spots are shown under five steps: I-AO off, phase AO on, I-AO SL1, phase AO SL1, and I-AO SL2. **(b)** Cross-sectional intensity profiles of the focal spot for each step in (a), compared with the ideal Airy disk (dashed line). The scale bar in (a) represents 30 μm for all normalised focus images.

To further validate the feasibility of our I-AO approaches, additional experiments were conducted, focusing on 1) errors introduced by a biological sample located between the pupil and focal plane, and 2) assessing the imaging of biomedical and material samples under a conventional wide-field microscope with intensity errors, both before and after applying I-AO. The results of both experiments confirm the effectiveness of the I-AO approach. Detailed procedures, results, and analysis are provided in **Supplementary Note 8**.

**Discussions**

In summary, we have extended the concept of conventional AO to include intensity error correction, prioritising intensity uniformity at the pupil plane while enabling the recovery of absolute intensity through a newly developed dual-loop feedback mechanism. We present two I-AO methods: a sensor-based approach and a sensorless method, both supported by proof-of-concept experiments. Notably, while conventional AO primarily addresses phase aberrations at the pupil plane, our method uniquely corrects for intensity errors at this plane, achieving properties including intensity uniformity at the pupil that traditional AO cannot attain.

In the sensor-based I-AO method, we focus on the pupil's intensity profile, using a camera as the primary sensor, while residual phase errors are corrected through conventional phase sensorless AO. However, conventional sensorless phase AO often misinterprets intensity errors as phase aberrations, making it ineffective in solely correcting intensity errors. In future developments, we plan to integrate a Shack-Hartmann wavefront sensor into the system for precise phase measurement to improve correction accuracy. For the sensorless I-AO method, we newly designed intensity Zernike modes, adapting conventional Zernike modes used in phase AO for flexibility. However, these intensity Zernike modes may not be the optimal set of modes for sensorless I-AO correction, as they are not orthogonal to each other and do not support superposition due to the energy related nature of intensity. They are primarily used here for conceptual demonstration purposes. In the future, zonal correction or other novel intensity modes may offer better alternatives than Zernike modes for enhancing correction accuracy and precision, providing a more effective sensorless I-AO procedure[34,36,37]. Future work could also explore the development of new image metrics and improved sensorless I-AO processes, possibly leveraging cutting-edge machine-learning techniques[2,38,39]. Additionally, integrating the I-AO toolkit with polarisation/vectorial AO and spatiotemporal control[39,40] will facilitate comprehensive field correction and particularly benefit complex optical systems and applications[41-44].

Overall, we have introduced a novel I-AO technique for correcting intensity errors. This advanced approach holds great potential as a versatile tool for applications in both scientific research and industry.

**Acknowledgments**

The authors would like to thank the support of Dr. Yun Zhang and Prof. Daniel Royston at the University of Oxford, St John's College (C.H.), and the Royal Society (URF\R1\241734) (C.H.).

**Conflict of interests**

Martin J. Booth serves as an Editor for the Journal. Stephen M. Morris and Chao He serves as the Guest Editors of Special Issue on Optics and Photonics at the University of Oxford for the Journal. No other author has reported any competing interests.